 \documentstyle[12pt]{crnart12}
\def\beq{\begin{equation}}
\def\eeq{\end{equation}}
\def\bea{\begin{eqnarray}}
\def\eea{\end{eqnarray}}
\def\bq{\begin{quote}}
\def\eq{\end{quote}}
\def\vev#1{\,\langle#1\rangle}

\begin{document}
\pagestyle{empty}
\begin{flushright}
{CERN-TH.6842/93}\\
{UMN-TH-1130/93}\\
{ACT-2/93}\\
{CTP-TAMU-11/93}
\end{flushright}
%\vspace*{5mm}
\begin{center}
{\bf FLIPPED ANGLES AND PHASES: A SYSTEMATIC STUDY}\\
\vspace*{1cm} {\bf John Ellis} \\
\vspace*{0.1cm}
{\it Theoretical Physics Division, CERN} \\
{\it CH - 1211 Geneva 23} \\
\vspace*{0.1cm}
and \\
\vspace*{0.1cm}
{\bf Jorge L. Lopez} and {\bf D.V. Nanopoulos}\\
\vspace*{0.1cm}
{\it Center for Theoretical
Physics, Department of Physics, Texas A\&M University}\\
{\it College Station, TX 77843-4242, USA}\\
{\it and}\\
{\it Astroparticle Physics Group, Houston Advanced Research Center
(HARC),}\\
 {\it The Woodlands, TX 77382, USA}\\
\vspace*{0.1cm}
and \\
\vspace*{0.1cm}
 {\bf Keith A. Olive} \\
\vspace{0.1cm}
{\it School of Physics and Astronomy, University of Minnesota} \\
{\it Minneapolis, MN 55455, USA}\\
\vspace*{1cm}
{\bf ABSTRACT} \\ \end{center}
%\vspace*{5mm}
{\rightskip=3pc
 \leftskip=3pc
 \tenrm
\baselineskip=12pt
\noindent
We discuss systematically the fermion mass and mixing matrices in
a generic \linebreak
field-theoretical flipped $SU(5)$ model, with particular
applications to neutrino and baryon number-changing physics. We
demonstrate that the different quark flavour branching ratios in
proton decay are related to the Cabibbo-Kobayashi-Maskawa angles,
whereas the lepton flavour branching ratios are undetermined.
The light neutrino mixing angles observable via oscillation
effects are related to the heavy conjugate (right-handed)
neutrino mass matrix, which also plays a key role in
cosmological baryogenesis. The ratios of neutrino and charged lepton
decay modes in baryon decay may also be related to neutrino
oscillation parameters. Plausible Ans\"atze for the generation
structure of coupling matrices motivate additional relations
between physical observables, and yield a satisfactory baryon
asymmetry.
\vglue 0.2cm}
\baselineskip=14pt

\vspace*{1cm}

\begin{flushleft}
CERN-TH.6842/93 \\
UMN-TH-1130/93 \\
{ACT-2/93}\\
{CTP-TAMU-11/93}\\
March 1993
\end{flushleft}
%\pagestyle{empty}
%\clearpage\mbox{}\clearpage
\voffset -1in
\textheight=23cm
\vskip2.0cm
\vfill\eject

\setcounter{page}{1}
\pagestyle{plain}
 {\newcommand{\la}{\mbox{\raisebox{-.6ex}{$\stackrel{<}{\sim}$}}}
{\newcommand{\ga}{\mbox{\raisebox{-.6ex}{$\stackrel{>}{\sim}$}}}
%\voffset -1in
%\textheight=20cm
%\vskip2.0cm

\section{Introduction}

One of the most welcoming
 avenues leading beyond the Standard Model is that leading to Grand
Unified Theories (GUTs).  By unifying the three known particle gauge
interactions within a simple group $G$, one may understand why baryon
number is conserved to a good approximation, but not perfectly, and why
neutrino masses may be very small, but non-zero.  In so doing, GUTs also
offer scenaria for cosmological baryogenesis and the nature of hot dark
matter.  Interest in GUTs has been further whetted by the close
consistency of the measured values of the SU(3), SU(2) and U(1) gauge
couplings with minimal supersymmetric GUTs \cite{susygut}, which
high-precision
 LEP
data have rendered even more impressive \cite{lep}.

With all these phenomenological motivations, it was natural that string
model-builders should seek to emulate GUTs.  However, it was soon
realized that there was a sizeable roadblock to deriving a GUT from
string:  gauge symmetry breaking and various other phenomenological
constraints
in GUTs generally require adjoint or larger Higgs
representations, and these are not obtainable using conventional
model-building technology based on $k = 1$ Kac-Moody currents on the
world-sheet \cite{string}.  Hence the revived
interest in an SU(5)$\times$U(1) GUT, flipped SU(5) \cite{barr,ant1},
which only
required 5- and 10-dimensional Higgs fields and could be derived from
string.  Flipped SU(5) is the closest homage string can pay to the
simple GUTs of old.

Naturally, there has been considerable discussion within flipped SU(5)
of the ``classic" new phenomena that motivated so much work on GUTs,
namely baryon decay \cite{bdecay} and neutrino masses
 \cite{ant1,ant2,leon,ln,eln,GLJV}.
  However, there has not been a
systematic investigation of all the fermion mass matrices and mixing
angles that enter into baryon decay branching ratios and neutrino
oscillations, and their possible relations to the
Cabibbo-Kobayashi-Maskawa (CKM) quark mixing matrix and cosmological
baryogenesis.  In this paper we seek to remedy this lack in the
literature, working in the general framework of minimal
field-theoretical flipped SU(5), supplemented at the end by some (to
us) plausible general hypotheses about mass matrices.

It may serve a useful purpose to recall first some important features
of minimal field-theoretical flipped SU(5) \cite{ant1}.
  The first is that baryon decay via
dimension-5 operators is very strongly suppressed, and also dimension-6
Higgs
exchanges are presumably negligible compared to the dimension-6 massive
vector boson exchange operators expected to dominate.  In minimal
SU(5), it was possible to relate the mixing angles in the corresponding
dimension-6 operators for proton decay to the CKM angles, modulo two
additional complex phases that would be difficult to measure \cite{egn}.
The
relation is not so direct in flipped SU(5), because the flipping of the
particle assignments within $\underline{\bar 5}$ and
$\underline{10}$ representations, as well as the assignment of
conjugate charged leptons to singlet representations, means the
branching ratios into different lepton species are independent of the
quark mixing angles.  It has recently been realized, on the other hand,
that the dominant mechanism for baryogenesis in flipped SU(5) may be
the decay of massive conjugate (``right-handed") neutrinos \cite{eno3},
 producing a
lepton asymmetry which is subsequently reprocessed \cite{fy1} into a
baryon
asymmetry by non-perturbative electroweak interactions \cite{krs1}.
This raises
 the
possibility that the cosmological baryon asymmetry might be related to
observable parameters in the light neutrino mass and mixing matrices,
which could also show up in baryon decay.

In this paper, we investigate these possibilities systematically,
starting from an adaptation in Section 2 of the analysis of ref.
\cite{egn}
to diagonalize the fermion mass matrices and determine the independent
unitary flavour rotation matrices.  Next, in Section 3 we apply these
results to derive interrelations between observables in conventional
weak decays, proton decay, neutrino oscillations and cosmological
baryogenesis.  Then, in Section~4 we postulate plausible Ans\"atze for
the fermion mass matrices which lead to certain additional quantitative
relations between the different flavour rotation matrices and hence
observable quantities.  Finally, in Section 5 we summarize our
conclusions and mention directions for future study.

\section{Flipped Mixing Matrices}

Let us first remind the reader of the various coupling matrices
that appear in the minimal field-theoretical version of
the flipped SU(5) model. The superpotential that
characterizes its Yukawa couplings is \cite{ant1}
\beq
W = \lambda_1^{ij}F_iF_jh + \lambda_2^{ij}F_i{\bar f}_j\bar h +
\lambda_3^{ij}\bar f_il^c_jh + \lambda_4 HHh +
 \lambda_5 \bar H \bar H \bar h
   \\
+ \lambda_6^{ia}F_i\bar
H \phi_a  +  {\lambda_7} h \bar h \phi_0  +  \mu_{ab}
\phi_a \phi_b
\label{sp}
\eeq
where the $F_i$, ${\bar f}_i$, ${l^c}_i$ (i = 1,2,3) are the
 three generations
of ${\bf 10}$, ${\bf {\bar 5}}$ and singlet representations of
 SU(5) that comprise the
light matter particles of the Standard Model, $H$ and ${\bar H}$
 are ${\bf 10}$ and
${\bf {\bar {10}}}$ Higgs representations,
 $h$ and ${\bar h}$ are ${\bf 5}$ and ${\bf {\bar 5}}$ Higgs
representations, and the $\phi_0,\phi_a$ (a = 1,2,3) are auxiliary
singlet
fields. The first 3 terms in the superpotential (\ref{sp}) give masses
to
the charge 2/3 quarks $u_i$, charge -1/3 quarks $d_i$ and charged
leptons $l_i$
respectively, the next two terms split the light Higgs doublets
from their heavy colour triplet partners in a natural way, the sixth
term provides a large element in the see-saw neutrino mass matrix,
the product $\lambda_7\vev{\phi_0}$ gives the traditional Higgs mixing
parameter, and the last term is the auxiliary singlet mass matrix
\cite{ant2}.

Our task in this section is to diagonalize the coupling
matrices $\lambda_{1,2,3,6}$ and the corresponding mass matrices
$m_{u,d,l}$, identifying in the process the needed unitary rotation
matrices whose observability we will explore in Section 3. Our
analysis of the diagonalizations is modelled on that made for
minimal SU(5) in reference \cite{egn}. We start by diagonalizing
the $\lambda_2$ coupling matrix with the unitary transformations
\beq
{\bar f'} = {\bar f}U_{u^c}^\dagger~~~~~~,~~~~~~F' = U_{u}^\dagger F
\label{t1}
\eeq
which yield
\beq
{\bar f} \lambda_2 F = {\bar f'} U_{u^c} \lambda_2 U_u F'
  =  {\bar f'} \lambda_2^D F'
{}~~~~~:~~\lambda_2^D =  U_{u^c} \lambda_2 U_u
\eeq
As a result, the $\lambda_1$ coupling term becomes
\beq
F^T \lambda_1 F = F'^T \lambda'_1 F'~~~~~:~~\lambda'_1 =
 U_u^T \lambda_1 U_u
\eeq
We now diagonalize this by a further unitary transformation
on the fields $F'$:
\beq
\widetilde F = U_4 F' ~~~~~:~~\lambda'_1 = U_4^T \lambda_1^D  U_4
\eeq
As in ref. \cite{egn}, it is convenient to separate phase factors
in the matrix $U_4$: denoting elements of this matrix by
$e^{i \eta_{ij}} u_{ij}$ where $\eta_{ij}$ and
$u_{ij}$ are both real, we decompose
\beq
U_4 = U_5 U U_6 ~~~~~:~~U_5 = diag(e^{i \eta_{i1}})~~~,
{}~U_6 = diag(e^{i \eta_{1j}}) e^{-i \eta_{11}}
\label{tt1}
\eeq
This leads to the representation
\beq
\lambda'_1 = U_6 U^T U_7 \lambda_1^D U U_6~~~~~:~~U_7 = U_5^2
\eeq
and we impose a phase convention: $det U_7$ = 1.
We can then absorb the $U_6$ phases into the fields
\beq
{\bar f''} = {\bar f'} U_6^*~~~~~,~~F'' = U_6 F'
\label{t2}
\eeq
leaving the diagonalized matrix ${\lambda_2}^D$ unchanged. In
this basis, the first term in equation (\ref{sp}) becomes
\beq
F''^T U^T U_7 \lambda_1^D U F''
\eeq
The charge -1/3 quark mass matrix may finally be cast in real
and diagonal form by the redefinitions
\begin{eqnarray}
F'''_{\alpha \beta} = U F''_{\alpha \beta}~~~~ \mbox{for $1
 \le \alpha,\beta \le 4$} \nonumber
\\
F^{IV}_{\alpha \beta} = U_7 F'''_{\alpha \beta} ~~~~ \mbox{for
 $1 \le \alpha,\beta \le 3$}
\label{t3}
\end{eqnarray}
of relevant components of the ${\bf 10}$ representations $F$.

In contrast to the conventional SU(5) case discussed in ref.
\cite{egn}, the charged lepton mass matrix is not directly related
to the quark mass matrices. Starting from the third term in the
superpotential (1) and making the above transformations of
 Eqs. (\ref{t1}, \ref{t2})
of the ${\bf 5}$ fields ${\bar f}$, we find
\beq
{\bar f} \lambda_3 l^c  =  {\bar f''}(U_6 U_{u^c} \lambda_3) l^c
\eeq
which we now diagonalize by the new transformations
\beq
{\bar f'''_5} = {\bar f''_5} U_l^\dagger~~~~~ :~~ {l^c}' =
 U_{l^c}^\dagger l^c
\eeq
leading to
\beq
{\bar f'''_5} \lambda_3^D {l^c}' ~~~~~:~~ \lambda_3^D = U_l
 U_6 U_{u^c} \lambda_3 U_{l^c}
\eeq
for the diagonalized mass matrix.

A further novel feature beyond the analysis of ref. \cite{egn}
is the diagonalization of the light and heavy neutrino mass
matrices. We start by diagonalizing the $\phi$ mass matrix:
\beq
\phi^T \mu \phi = \phi'^T \mu^D \phi' ~~~~~ :~~ \phi'
 = U_\phi^\dagger \phi
 ~~~~,~\mu^D = U_\phi^T \mu U_\phi
\eeq
which leads to the representation
\beq
F^T \lambda_6 \phi = F''^T \lambda'_6 \phi'~~~~~~ :~~ \lambda'_6
 = U_6^* U_u^T \lambda_6 U_\phi
\eeq
The $\nu^c = F_{45}$ mass matrix is of the see-saw form in the
($F_{45}'', \phi'$) basis, and so can be written in the form
\beq
(F''_{45}  + \dots)^T m'_{\nu^c} (F''_{45}  + \dots)
{}~~~~~ :~~ m'_{\nu^c} = {\lambda'_6}^T ({\mu^D})^{-1} \lambda'_6
\,\vev{\bar V}^2,
\eeq
where $\vev{\bar V}=\vev{\overline{\bf10}}$.
We recall that the transformations (\ref{t3}) applied to the coloured
components of $F$ in order to diagonalize the quark mass matrices
have not been applied to the $\nu^{c} = F_{45}$ components. We now
define
\beq
F'''_{45} = U_{\nu^c}^\dagger F''_{45}
\eeq
in terms of which the $\nu^c$ mass matrix is diagonal:
\begin{eqnarray}
{F'''_{45}}^T m_{\nu^c}^D F'''_{45} ~~~~~: ~~m_{\nu^c}^D =
U_{\nu^c}^T m'_{\nu^c} U_{\nu^c} = U_{\nu^c}^T
 {\lambda'_6}^T {(\mu^D)}^{-1} \lambda'_6 \,\vev{\bar V}^2\,U_{\nu^c}  =
\nonumber \\
 U_{\nu^c}^T (U_\phi^T \lambda_6^T U_u U_6^*) (U_\phi^\dagger
\mu^{-1} U_\phi^*) (U_6^* U_u^T \lambda_6 U_\phi)\,\vev{\bar V}^2\,
U_{\nu^c}
\label{mess}
\end{eqnarray}
The light neutrino mass matrix is also of the see-saw form, this time
in the ($\nu = {\bar f}_4$, $F_{45}$) basis. We diagonalize it in terms
of the diagonalized $\nu^c$ masses $m^D_{\nu^c}$ (\ref{mess}). In terms
of
the unitary transformations already made:
\beq
{\bar f_4} \lambda_2 F_{45} = {\bar f''_4} \lambda_2^D U_{\nu^c}
F'''_{45}
\eeq
yielding
\beq
{\bar f''_4} m_\nu {{\bar f''}_4}{}^T ~~~~~:~~ m_\nu =
 \lambda_2^D U_{\nu^c} ({m_{\nu^c}^D})^{-1} U_{\nu^c}^T \lambda_2^D
\label{twenty}
\eeq
which we must diagonalize by the further transformation
\beq
{\bar f'''_4} = {\bar f''_4} U_\nu^\dagger ~~~~~:~~
m_\nu^D = U_\nu [\lambda_2^D U_{\nu^c}
({m_{\nu^c}^D})^{-1} U_{\nu^c}^T \lambda_2^D] U_\nu^\dagger
\label{twentyone}
\eeq
for the light neutrino mass eigenstates.

After this profusion (confusion?) of diagonalizations and unitary
matrices, we take mercy on the reader by summarizing the final
mass eigenstates (represented by suffices $F$):
\begin{eqnarray}
u_F = (U_6 U_u^\dagger) F_{1 \le \alpha \le 3,5}
{}~~~~~~,~~~~~~~u_F^c  = {\bar f_{1 \le \alpha \le 3}}
 (U_{u^c}^\dagger  U_6^*) \label{ul} \\
d_F =  (U U_6 U_u^\dagger) F_{1 \le \alpha \le 3,4}
{}~~~~~,~~d_F^c =  (U_7 U U_6 U_u^\dagger)
F_{1 \le \alpha,\beta \le 3}   \label{dl} \\
l_F = {\bar f_5} (U_{u^c}^\dagger  U_6^* U_l^\dagger)
{}~~~~~,~~~~~~~~~~~~~~~~~~~~~~l_F^c = U_{l^c}^\dagger l^c  \label{L} \\
\nu_F  =  {\bar f_4} (U_{u^c}^\dagger  U_6^* U_\nu^\dagger)
{}~~~~~,~~~~~~~~~~\nu_F^c = (U_{\nu^c}^\dagger
 U_6 U_u^\dagger) F_{45}  \label{nu}
\end{eqnarray}

\section{Specific Processes}

As we mentioned earlier, among the motivations for
 Grand Unified Theories
is the possibility for baryogenesis and small neutrino masses.
One of the remarkable features of the flipped SU(5) model
 is its ability
to provide cosmologically-interesting neutrino masses
 (i.e., for $\nu_\tau$),
while at the same time allowing for observable
Mikheyev-Smirnov-Wolfenstein (MSW) \cite{msw}
neutrino oscillations
 and baryogenesis
via leptogenesis and subsequent
sphaleron reprocessing. All three of these
 highly desirable
features are related to the {\it same} neutrino mass matrix
\cite{eln,eno3}.
Given the complete decomposition of mass eigenstates in the
 flipped SU(5) model above,
we are now in a position to examine in detail the
 interrelationships between the
various mixing matrices and phases of interest.
  In particular, we will be interested
in identifying the mixing matrix for the MSW mixing
 and the origin of the necessary
CP violating phase for the production of the cosmic
 lepton asymmetry. We will also
identify the Cabibbo-Kobayashi-Maskawa (CKM)
matrix and its role, as well as
those of other
matrices of interest
 in proton decay processes.

We begin with the MSW mixing matrix.  MSW solar neutrino
 mixing will result if there
is a mismatch between the neutrino
states produced
in the charged current interactions and the neutrino
 mass eigenstates.
Therefore, we can define the MSW mixing matrix by
\beq
 l {\bar \nu} =   l_F U_{MSW}{\bar \nu_F}
\eeq
Given the previous results for the mass eigenstates
 $l_F$ (\ref{L}) and $\nu_F$
 (\ref{nu}) we find quite simply
\beq
U_{MSW} = U_l U_{\nu}^{\dagger}
\label{UMSW}
\eeq
Similarly, the CKM matrix is the charged current
 mismatch in the quark sector,
\beq
{\bar u} d = {\bar u_F} U_{CKM} d_F
\eeq
Using now the expressions (\ref{ul}) and (\ref{dl})
 for the mass eigenstates $u_F$ and
$d_F$ we can write
\beq
U_{CKM} = U^{\dagger}
\label{UCKM}
\eeq
$A~priori$, we see no relationship between
 the quark and neutrino mixing
matrices.

Next we study the effective Lagrangian for baryon decay in
the flipped SU(5) model, which has been discussed previously
in ref. \cite{EHKN}. As has already been recalled, the dominant
mechanism for baryon decay is expected to be dimension-6 vector
boson exchange. Using the analysis of the previous section, as
was done for conventional SU(5) in ref. \cite{egn}, we find
the following massive vector boson couplings:
\begin{eqnarray}
{\cal L}_X = \frac{g}{\sqrt{2}} {{X_i^-}_\mu} [\epsilon^{ijk}
{\bar d^c}_{k_F} U_7 \gamma^\mu P_L d_{j_F} + {\bar u}_{i_F}
 \gamma^\mu P_R \nu^c ]   +  h.c.  \nonumber \\
{\cal L}_Y = \frac{g}{\sqrt{2}} {{Y_i^-}_\mu} [\epsilon^{ijk}
{\bar d^c}_{k_F} U_7 U \gamma^\mu P_L u_{j_F} + {\bar u}_{i_F}
 \gamma^\mu P_R l^c ]   +  h.c.
\end{eqnarray}
where we have worked in the mass eigenstate basis for the
quarks, denoted by subscripts $F$, the lepton fields are in the
${\bar f}''$  basis of the previous section, the colour indices
$(i,j,k)$ are noted explicitly, and we have indicated the handednesses
of all fermion fields by using the projection operators $P_L$ and $P_R$.
 The exchanges of $X$ and $Y$ bosons
(assumed as usual to have indistinguishable masses) therefore give
rise to the following effective Lagrangian for baryon decay:
\begin{eqnarray}
{\cal L}_{\Delta B \ne 0} & =  & \frac{g^2}{2 M_X^2} \left[
(\epsilon^{ijk}
{\bar d^c}_{k_F} U_7 \gamma^\mu P_L d_{j_F}) (u_{i_F}
 \gamma_\mu P_L \nu)  +  h.c. \right. \nonumber \\
 & + & \left. (\epsilon^{ijk}
{\bar d^c}_{k_F} U_7 U \gamma^\mu P_L u_{j_F}) (u_{i_F}
 \gamma_\mu P_L l) +  h.c. \right]
\end{eqnarray}
As in the case of conventional SU(5), we see from this
expression that two additional $CP$-violating phases appear in
the quark parts of the $\Delta B \ne 0$ operators [{\it i.e.}, from
$U_7=U^2_5$
in Eq. (\ref{tt1}) and $\det U_7=1$], beyond
the single phase in the CKM matrix: these are in principle
measurable via loop diagrams. Since the only quarks of relevance
for baryon decay at the tree level are $u$, $d$ and $s$, and not
more than one of the latter, we can write the relevant parts
of the $\Delta B \ne 0$ Lagrangian (above) as:
\begin{eqnarray}
{\bar {\cal L}}_{\Delta B \ne 0} & =  & \frac{g^2}{2 M_X^2}
 \left[ (\epsilon^{ijk}
{\bar d^c}_{k} e^{2 i \eta_{11}} \gamma^\mu P_L d_{j}) (u_{i}
 \gamma_\mu P_L \nu_1)  +  h.c. \right. \nonumber \\
 & + & \left. (\epsilon^{ijk}
({\bar d^c}_{k} e^{2 i \eta_{11}} \cos \theta_c
 + {\bar s^c}_{k} e^{2 i \eta_{21}} \sin \theta_c)
\gamma^\mu P_L u_{j}) (u_{i}
 \gamma_\mu P_L l_1) +  h.c. \right]
\label{32}
\end{eqnarray}
where we recall that $\nu_L$ and $l_L$ are not in the mass eigenstate
basis, so that
\beq
\nu_L =\nu_F U_\nu ~~~~~ , ~~l_L =l_F U_l
\label{33}
\eeq
The phase factors in (\ref{32}) are not measurable, nor can
one distinguish between the different neutrino flavours, so
the following are the predictions that can be made on the
basis of (\ref{32}):
\beq
\frac {\Gamma (B \rightarrow (\Sigma \nu) + X|_{strange})}
{\Gamma (B \rightarrow (\Sigma \nu) + X|_{non-strange})} = 0~~, ~~
\frac {\Gamma (B \rightarrow l^+ + X|_{strange})}
{\Gamma (B \rightarrow l^+ + X|_{non-strange})} = \tan ^2 \theta_c
\approx {1\over 20}
\eeq
Comparing decays to neutrinos and charged leptons requires
knowledge of specific hadronic matrix elements. Following
ref. \cite{EHKN}, we find
\begin{eqnarray}
\Gamma(p \rightarrow e^+ \pi^o) = \frac{\cos ^2 \theta_c}{2}
 |U_{l_{11}}|^2
\Gamma(p \rightarrow {\bar \nu} \pi^+) = \cos ^2 \theta_c
|U_{l_{11}}|^2
\Gamma(n \rightarrow {\bar \nu} \pi^o) \nonumber \\
\Gamma(n \rightarrow e^+ \pi^-) = 2
 \Gamma(p \rightarrow e^+ \pi^o) ~~,~~
\Gamma(n \rightarrow \mu^+ \pi^-) = 2
\Gamma(p \rightarrow \mu^+ \pi^o) \nonumber \\
\Gamma(p \rightarrow \mu^+ \pi^o) = \frac{\cos ^2 \theta_c}{2}
 |U_{l_{12}}|^2
\Gamma(p \rightarrow {\bar \nu} \pi^+) = \cos ^2 \theta_c
|U_{l_{12}}|^2
\Gamma(n \rightarrow {\bar \nu} \pi^o)
\label{gammas}
\end{eqnarray}
Thus it is possible in principle to correlate baryon decay branching
ratios with the CKM angles, and to measure elements of the
charged-lepton mixing matrix $U_l$, though not of the neutrino
mixing matrix $U_{\nu}$. Moreover, we note that the decay
branching ratios above are characteristically different from
those in conventional SU(5) \cite{EHKN}. However, it lies beyond the
scope of
this paper to calculate the total baryon decay rate in flipped SU(5).

We now examine the mixing
matrix and the necessary CP-violating
phase which
can provide for a net lepton asymmetry, and subsequently
 a net baryon asymmetry via sphaleron reprocessing \cite{fy1}.
As described in Ref. \cite{eno3}, the lepton asymmetry is produced
 by the out-of-equilibrium
decay of the mass eigenstate  $\nu^c_F = F'''_{45}$ via the mass
 term ${\bar f_{4,5}} \lambda_2 F_{45}$.
In terms of mass eigenstates the neutrino mass term becomes
\beq
{\bar f'''_{4,5}} U_{\nu,l}\, \lambda_2^D U_{\nu^c} F'''_{45}
\eeq
It is then natural to introduce the lepton-number-violating coupling
\beq
\lambda_L \equiv U_{\nu,l}\, \lambda_2^D U_{\nu^c}
\eeq
We expect the dominant contribution to the lepton (baryon)
asymmetry to be that due to decays of the lightest $\nu^c$ mass
eigenstates, which we expect to be that associated with $\nu_{e,\mu}$
and call $\nu_{1,2}$.
The CP asymmetry in the decay of $\nu_1^c$ into 2nd and 3rd generation
particles is given by
\beq
\epsilon_1 =  \frac{1}{2 \pi (\lambda_L^\dagger
\lambda_L)_{11}}
 \sum_j \left({\rm Im}[(\lambda_L^{\dagger} \lambda_L)_{1j}]^2 \right)
g(M_j^2/M_1^2)
\label{cp}
\eeq
where
\beq
g(x) = 4 \sqrt{x} \ln \frac{1 + x}{x} ~~~~~(\simeq \frac{4}
{\sqrt{x}} ~~~~; x \gg 1)
\eeq
where (\ref{cp}) is the supersymmetric expression for
 the CP asymmetry \cite{cdo2}. Analogous expressions
can be written for the corresponding asymmetries
$\epsilon_{2,3}$ in the decays
of $\nu_{2,3}^c$.
Using the definition (31) of $\lambda_L$ we find
\beq
\lambda_L^{\dagger} \lambda_L = U_{\nu^c}^\dagger
 (\lambda_2^D)^2 U_{\nu^c}
\label{bb}
\eeq
This depends only on $U_{\nu^c}$ and, what is more,
 the CP-violating phase we
are interested in is $a$ $priori$ unrelated to
the CKM phase or MSW mixing.
In general, we could expect $(\lambda_L^{\dagger}
 \lambda_L)_{11} \sim ({\lambda_2^D}_{33})^2$
(the largest entry in $\lambda_2^D$), and if it were the case that
 $M_1 \sim M_2 \ll M_3$
we would estimate
\beq
\epsilon \simeq \frac{2 \ln 2 }{\pi}
 |{\lambda_2^D}_{33}|^2 \delta
\eeq
where $\delta$ is the phase associated with
 the imaginary part of
 $(\lambda_L^{\dagger} \lambda_L)_{12}$ in (\ref{bb}).
  (This is slightly larger
than our previous estimate \cite{eno3}.)
 A satisfactory baryon asymmetry would result
for $\delta \ga 10^{-2}$. In the next section we will revise this
generic
estimate in the light of a specific Ansatz for neutrino masses.

 \section{Phenomenological Ans\"atze}
In this section we propose some plausible forms for the various matrices
appearing
above, and obtain correlations among the various observables of
interest.
Clearly, the unitary matrices $U$ are not expected to be equal to the
identity matrix. However, experience with the CKM and MSW mixing
matrices
indicates
that they probably should not differ too much from unity either. We
therefore
write
\beq
U_i={\bf1}+R_i,
\eeq
for all matrices $U_i$ defined above, with
\beq
R_i=\pmatrix{0&\theta^i_{12}&0\cr -\theta^{*i}_{12}&0&\theta^i_{23}\cr
0&-\theta^{*i}_{23}&0\cr},
\eeq
such that $U_iU^\dagger_i={\bf1}$ through order $\theta$, since
$R^\dagger_i=-R_i$.
By analogy with the CKM matrix, we have neglected the far off-diagonal
entries in the matrices $R_i$.

For the MSW matrix $U_{MSW}=U_l U^\dagger_\nu$ we have
$R_{MSW}=R_l+R^\dagger_\nu=R_l-R_\nu$ and
\beq
\theta_{e\mu}=\theta^l_{12}-\theta^\nu_{12},
\qquad\theta_{\mu\tau}=\theta^l_{23}
 -\theta^\nu_{23},
\label{angles}
\eeq
where $\theta_{e\mu,\mu\tau}$ are the usual MSW mixing angles. It is
interesting to note
that the baryon decay branching fractions in Eq. (\ref{gammas}) depend
in this
approximation on
\beq
U_{l_{11}}=1,\qquad U_{l_{12}}=\theta^l_{12},
\eeq
giving
\beq
\Gamma(p\to e^+\pi^0)={1\over2}\cos^2\theta_c\Gamma(p\to\bar\nu\pi^+)=
\cos^2\theta_c\Gamma(n\to\bar\nu\pi^0).
\eeq
Note that our general Ans\"atze about mixing angles indicate that the
baryon
decay branching fractions into muons should be suppressed relative to
those
into electrons. Also, $R_\nu$ remains undetermined this way since only
the sum over all branching fractions into neutrino final states can be
observed.  On the other hand, Eqs. (\ref{twenty},\ref{twentyone}) allow
us to
obtain $U_\nu$ once $U_{\nu^c}$ and $m^D_{\nu^c}$ are given. Since the
latter two appear in the baryogenesis parameter $\epsilon$ (Eqs.
\ref{cp},\ref{bb}), it is
interesting to correlate all these parameters.

Starting from Eq. (\ref{twentyone}) and writing $U_\nu={\bf1+}R_\nu$ and
$U_{\nu^c}={\bf1}+R_{\nu^c}$
we obtain
\beq
m_{\nu_i}=(\lambda^D_2)_i(m^D_{\nu^c})^{-1}_i(\lambda^D_2)_i,
\label{masses}
\eeq
which leads to the previously advocated phenomenologically interesting
neutrino
mass ratios \cite{eln}.
Moreover, from $(m^D_\nu)_{ij}=0$ for $i\not=j$ we find $R_\nu$ in terms
of $R_{\nu^c}$,
\beq
(R_\nu)_{ij}={-1\over m_{\nu_j}-m_{\nu_i}}(\lambda^D_2)_i(\lambda^D_2)_j
\left[(R_{\nu^c})_{ij}{1\over M_j}-(R^*_{\nu^c})_{ij}{1\over
M_i}\right],\eeq
where $M_i\equiv(m^D_{\nu^c})_{ii}$. In particular
\beq
\theta^\nu_{12}={-\lambda_u\lambda_c\over m_{\nu_\mu}-m_{\nu_e}}
\left[\theta^{\nu^c}_{12}{1\over M_2}-\theta^{*\nu^c}_{12}{1\over
M_1}\right]
 \to{\lambda_u\over\lambda_c}\theta^{*\nu^c}_{12}{M_2\over M_1}.\eeq
The last expression holds in the limit $M_2\gg M_1$. Also,
\beq
\theta^\nu_{23}={-\lambda_c\lambda_t\over m_{\nu_\tau}-m_{\nu_\mu}}
\left[\theta^{\nu^c}_{23}{1\over M_3}-\theta^{*\nu^c}_{23}{1\over
M_2}\right]
\to{\lambda_c\over\lambda_t}\,\theta^{*\nu^c}_{23}{M_3\over M_2},
\eeq
with the final expression valid in the limit $M_3\gg M_2$.

As is plausible for any matrix with a hierarchy of eigenvalues,
such as a see-saw mass matrix, we make the
Ansatz
\beq
\theta^{\nu^c}_{12}\simeq \sqrt{M_1\over M_2},
\qquad\theta^{\nu^c}_{23}\simeq \sqrt{M_2\over M_3},
\eeq
which gives
\beq
\theta^\nu_{12}\simeq{\lambda_u\over\lambda_c}\sqrt{M_2\over M_1},\qquad
\theta^\nu_{23}\simeq{\lambda_c\over\lambda_t}\sqrt{M_3\over M_2}.
\label{newpred}
\eeq
If for the moment we neglect the $\theta^l_{12,23}$ contributions to the
neutrino mixing angles [see Eq. (\ref{angles})], the prediction for
$\theta_{\mu\tau}$ in Eq. (\ref{newpred}) is {\it exactly} what was
proposed in
Ref. \cite{eln} on purely phenomenological grounds. This relation (with
$M_3/M_2\sim10$) and the ratios of neutrino masses in Eq. (\ref{masses})
have
been shown \cite{eln} to lead to interestingly observable
$\nu_\mu-\nu_\tau$
oscillations and a $\tau$ neutrino mass large enough to provide an
interesting
amount of astrophysical hot dark matter. The other relation in Eq.
(\ref{newpred}) should reproduce the present fits to the solar neutrino
data
based on the MSW mechanism. These require
$\theta_{e\mu}=(3.2-6.1)\times10^{-2}$
\cite{Gallex} and therefore $M_2/M_1=64-225$. (Note: the present
expression for
 $\theta_{e\mu}$ differs from that in Ref. \cite{eln}.)

%Since $\lambda_u/\lambda_c<10^{-2}$, it is plausible to assume that
%$|\theta^\nu_{12}|
%\ll\theta^l_{12}$. In this case Eq. \ref{angles} implies
%\beq
%\sin^2(2\theta_{e\mu})=
%{8\over\cos^2\theta_c}{\Gamma(p\to\mu^+\pi^0)\over\Gamma(p\to\bar\nu\pi
%={4\over\cos^2\theta_c}{\Gamma(p\to\mu^+\pi^0)\over\Gamma(n\to\bar\nu\p
%\eeq
%which correlates neutrino mixing with baryon decay.

Turning now to the baryogenesis parameter $\epsilon$, from Eq.
(\ref{bb})
we have (to order $\theta$)
\beq
(\lambda^\dagger_L\lambda_L)_{ij}=(\lambda^D_2)^2_{ij}\delta_{ij}
+(R_{\nu^c})_{ij}\left[(\lambda^D_2)^2_i-(\lambda^D_2)^2_j\right].\eeq
Thus ${\rm Im}\,(\lambda^\dagger_L\lambda_L)^2_{ii}={\rm
Im}\,(\lambda^\dagger_L\lambda_L)^2_{13}=0$, and
\begin{eqnarray}
{\rm Im}\,(\lambda^\dagger_L\lambda_L)^2_{12} \simeq
\lambda^4_c|\theta^{\nu^c}_{12}|^2\cos2\phi_{12} \nonumber \\
{\rm Im}\,(\lambda^\dagger_L\lambda_L)^2_{23} \simeq
\lambda^4_t|\theta^{\nu^c}_{23}|^2\cos2\phi_{23}
\end{eqnarray}
where $\phi_{ij} = \arg (\theta_{ij}^{\nu^c})$.
Analogously, the denominator factors are given by
\begin{eqnarray}
%% FOLLOWING LINE CANNOT BE BROKEN BEFORE 80 CHAR
%% FOLLOWING LINE CANNOT BE BROKEN BEFORE 80 CHAR
%% FOLLOWING LINE CANNOT BE BROKEN BEFORE 80 CHAR
(\lambda^\dagger_L\lambda_L)_{11} & \simeq &
\lambda^2_u+\lambda^2_c|\theta^{\nu^c}_{12}|^2 \nonumber \\
(\lambda^\dagger_L\lambda_L)_{22} & \simeq &
\lambda^2_c+\lambda^2_t|\theta^{\nu^c}_{23}|^2 \nonumber \\
(\lambda^\dagger_L\lambda_L)_{33} & \simeq &
\lambda^2_t
\end{eqnarray}
We finally get
\begin{eqnarray}
\epsilon_1 & = & {2\over\pi}\lambda^2_c{M_1\over M_2}\delta_{12}
\nonumber \\
\epsilon_2 & = & {2\over\pi}\lambda^2_t{M_2\over M_3}\delta_{23}
\nonumber \\
\epsilon_3 & = & {4\over\pi}\lambda^2_t{M_2\over M_3}\ln{M_3\over M_2}
|\theta^{\nu^c}_{23}|^2
\delta_{23}
\end{eqnarray}
where
\beq
\delta_{12}=\cos2\phi_{12},\qquad
\delta_{23}=\cos2\phi_{23}
\eeq
For the decay of the first and lightest
generation $\nu_1^c$, this
result is a factor of $(\lambda_c/\lambda_t)^2 ({M_1 \over {\ln 2
M_2}})
\sim10^{-6}$ smaller than
that obtained in the generic case in Eq. (41)
above. However, we see that for the decay
of the second generation $\nu_2^c$ the result is only suppressed by
$({M_2 \over {\ln 2 M_3}}) \sim 7$, and is therefore large enough
$a$ $priori$ to produce a satisfactory lepton, and hence baryon,
asymmetry.

Normally, it is the lightest generation which produces a net
asymmetry \cite{fot3}. This is because the lightest generation
will typically wipe out any prior asymmetries (e.g., via inverse decays)
 before producing its own.
If we considered a standard out-of-equilibrium decay mechanism,
we would have
to require that both first- and second- generation $\nu^c$'s be far
out-of-equilibrium
at the time of their decay. This would impose the restrictions
$(\lambda^\dagger_L\lambda_L)_{11,22} \la 10^3 M_{1,2} / M_P$,
conditions
which would be difficult to satisfy given our Ans\"atze.

However, invoking a mechanism first proposed in Ref. \cite{NOST} we
assume here
that the $\nu^c$'s are produced in the decay of the inflaton, $\eta$,
subsequent
to inflation. Then $\nu^c$ decays
occur immediately and out-of-equilibrium at $T \ll M_{\nu^c}$,
and no destruction of an asymmetry can occur.  Thus our only requirement
is that $M_{1,2} < m_\eta$. This is satisfied for $m_\eta \sim few
\times
 10^{11} GeV$
as inferred from the COBE result \cite{cobe} on density fluctuations
 \cite{cdo2,eno3}.
Our final lepton and hence baryon asymmetry now becomes
\beq
\frac{n_B}{n_\gamma} \sim \frac{n_L}{n_\gamma} \simeq \frac{1}{\Delta}
\left( \frac{m_\eta}{M_P} \right)^{1/2}
(\epsilon_1 + \epsilon_2) \sim  10^{-7} \frac{M_2}{M_3} \delta_{23} \sim
10^{-8} \delta_{23}
\label{58}
\eeq
where $\Delta \sim 10^{-3}$ is an entropy dilution factor which accounts
for the
entropy produced during breaking of $SU(5) \times U(1)$ \cite{ent,eno3}.
This is encouragingly large: recall the
phase $\delta_{23}$
is not related to the
standard (CKM)
source of CP violation, and is not expected to be particularly
suppressed. Note that there could be extra sources of entropy which
would
further suppress the estimate (\ref{58}), so $\delta_{23}$ may not need
to be
very small.

\section{Summary and Outlook}

We have made in this paper a systematic study of the mass matrices
and mixing angles in the minimal field-theoretical version of
flipped SU(5). We have identified the mixing angles that appear
in observable processes such as neutrino oscillations and proton
decay, as well as the CKM angles in the charged electroweak
interactions, and related them as far as possible in a
independent way. We have also discussed a scenario for producing
the baryon asymmetry via the out-of-equilibrium decays of heavy
singlet conjugate neutrinos, which produce a lepton asymmetry that
is subsequently reprocessed into a baryon asymmetry by sphaleron
transitions. Additional plausible Ans\"atze for the mixing
angles lead to a number of further relations, and a satisfactory
baryon asymmetry.

We were motivated to study flipped SU(5) because it is the only
GUT that can be derived from string theory, and it was precisely
the baryon- and lepton-number-violating processes discussed above
that motivated the derivation of a GUT from string. However, the
minimal field-theoretical flipped SU(5) model analyzed above is not
what one obtains from string theory. On the one hand, flipped
SU(5) models derived from string contain more states, thus
complicating the analysis, but on the other hand they can cast
light on plausible forms for the mass matrices and mixing angles.
Therefore it would be desirable to complement this general
analysis with some more model-dependent studies.

We believe that flipped SU(5) can give us many insights into the
exciting new era of massive neutrino physics that solar neutrino
experiments and models of the formation of large-scale astrophysical
structures
suggest may be opening up before us.

\vspace*{1cm}
\noindent {\bf Acknowledgements}

 The work of JLL has been supported by an SSC Fellowship.
The work of DVN was supported in part by DOE grant
DE-FG05-91-ER-40633 and by a grant from Conoco Inc. The work
of KAO was supported in part by DOE grant DE-AC02-83ER-40105, and by
a Presidential Young Investigator Award. JLL would like to thank the
CERN Theory Division for its kind hospitality while part of this work
was carried out.

\end{document}